\documentclass[aps,prx,twocolumn,showpacs,superscriptaddress,groupedaddress,floatfix,10pt,amsmath,amssymb,footinbib]{revtex4-2}
\usepackage{graphicx}
\usepackage{bm,bbm}
\usepackage[urlcolor=blue]{hyperref}
\usepackage{amssymb} 

\newcommand{\E}[1]{\left\langle{ #1}\right\rangle}
\newcommand{\Es}[1]{\E{ #1}_{\rm s}}

\newcommand{\rmd}{d}
\newcommand{\rme}{\mathrm{e}}

\newcommand{\f}[1]{\mathbf{#1}}
\newcommand{\x}{\f x}
\newcommand{\y}{\f y}
\newcommand{\z}{\f z}

\newcommand{\abs}[1]{\left\lvert #1 \right\rvert}

\newcommand{\bsig}{\boldsymbol{\sigma}}

\newcommand{\bj}{\hat{\f j}}
\newcommand{\ps}{p_{\rm s}}
\newcommand{\js}{{\f j}_{\rm s}}

\newcommand{\revj}{{\hat{\f j}}^\ddag}
\newcommand{\integrals}{\hat{\mathcal{I}}}

\usepackage{color}\usepackage{xcolor}
\colorlet{mylinkcolor}{blue!66!black!80}
\newcommand{\blue}[1]{{\color{mylinkcolor}#1}}
\hypersetup{colorlinks=true,allcolors=mylinkcolor}

\renewcommand{\blue}[1]{{#1}}

\newcommand{\finalchange}[1]{{\color{black}#1}}

\begin{document}
\title{Mathematical, Thermodynamical, and Experimental Necessity
    for Coarse Graining Empirical Densities and Currents
  in Continuous Space}
\author{Cai Dieball}
\author{Alja\v{z} Godec} 
\email{agodec@mpinat.mpg.de}
\affiliation{Mathematical bioPhysics Group, Max Planck Institute for Multidisciplinary Sciences, Am Fa\ss berg 11, 37077 G\"ottingen}
\begin{abstract}
We present general results on fluctuations and spatial correlations of 
the coarse-grained empirical density and current of Markovian
diffusion in equilibrium or non-equilibrium steady states on all
time scales.~We unravel a deep connection between current
fluctuations and generalized time-reversal symmetry,
providing new insight into time-averaged observables.~We
highlight the essential role of coarse graining in space from
mathematical, thermodynamical, and experimental points of view.~Spatial coarse graining is required to uncover salient features of currents that break detailed balance, and a thermodynamically "optimal" coarse graining ensures the most 
  precise inference of dissipation.~Defined without coarse
  graining, the fluctuations of empirical density and current are
  proven to diverge on all time scales in dimensions higher than
  one, which has far-reaching consequences for \finalchange{the central-limit regime}
  in continuous space.~We apply the results to 
examples of
irreversible diffusion.~Our findings provide new intuition about
time-averaged observables
and allow for a more efficient analysis of single-molecule
experiments.
\end{abstract}
\maketitle

\begin{figure}[ht!!]
\includegraphics[width=.45\textwidth]{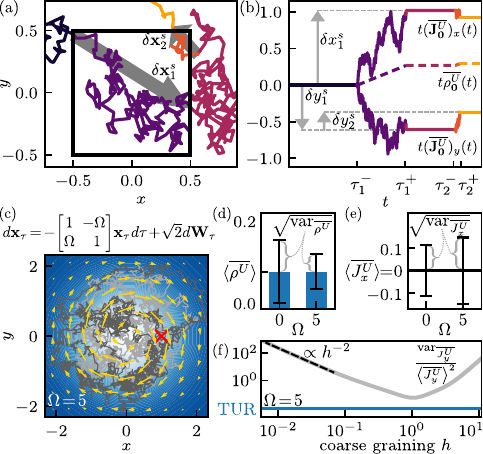}
\caption{(a)~Diffusive trajectory traversing an observation window \blue{$U^h_{\f 0}(x,y)=1$ if $\abs{x},\abs{y}\le 1/2$ and $U^h_{\f 0}(x,y)=0$ otherwise,} with time running from dark
  to bright.~Arrows denote contributions $\delta\x_i^s=(\delta
  x^s_i,\delta y^s_i)$ of the two sojourns in $U^h_{\f 0}$ between
  times $\tau_i^-$ and  $\tau_i^+$ \blue{(see Eq.~\eqref{current as sum of crossings} in Appendix~I)}.~(b)~Corresponding $t\overline{\rho^U_{\f 0}}(t)$ and
  components of $t\overline{\f J^U_{\f 0}}(t)$ from
  Eq.~\eqref{def_current} as functions of $t$.~(c)~Two
    trajectories $(\mathbf{x}_{\tau})$ (gray lines) of length $t=5$ in
    confined rotational flow with 
   $\Omega=5$ (arrows depict the steady-state current
    $\js$).~The red cross is the reference point $\x_R=(1,0)$
    considered in (d-f).~Coarse-grained density~(d) and
    $x$-current (e)~for a Gaussian window $U^h_{\x_R}$ with
    $h=0.3$. 
    Fluctuations of $\overline{\rho^U_\x}$ and
  $\overline{J^U_{x,y}}\equiv(\overline{\f J^U_{\x_R}})_{x,y}$ encode violations of
    detailed balance even where $\overline{J_{x}^U}$
    vanishes.~(f)~Squared relative error of $\overline{J_{y}^U}$ for 
  $U^h_{\x_R}$ as a function of 
  $h$ (gray) bounded by the thermodynamic
uncertainty relation (TUR; blue).
A variance diverging as $h^{-2}$ (dashed)
as
$h\!\to\!0$ and 
vanishing mean for
$h\!\gg\!1$
allow for intermediate
$h$ optimizing the TUR-bound and thus 
the inferred 
dissipation.}
\label{Fg1}
\end{figure}

Single-molecule experiments
\cite{Gladrow2016PRL,Gnesotto2018RPP,Ritort2006JPCM,Greenleaf2007ARBBS,Moffitt2008ARB}
probe equilibrium and non-equilibrium (i.e.~detailed balance
violating)
processes during
relaxation
\cite{Vaikuntanathan2009EEL,Maes2011PRL,Qian2013JMP,Maes2017PRL,Shiraishi2019PRL,Lapolla2020PRL,Koyuk2020PRL}
or in 
steady states
\cite{Jiang2004,Maes2008PA,Maes2008EEL,Gingrich2016PRL,Barato2015PRL,Dechant2018JSMTE,Seifert2010EEL,Rondoni2021PRE,Rondoni2016JSP}
on the level of individual 
trajectories.\ These are typically analyzed 
by averaging along individual
realizations yielding random quantities with nontrivial
statistics \cite{Burov2011PCCP,Lapolla2020PRR}.~Time-averaged
  observables, in particular 
  generalized currents,
  are central to stochastic thermodynamics \cite{Qian_PRE,Horowitz2019NP,Gingrich2016PRL,Gingrich2017JPAMT,Dechant2021PRR,Dechant2021PRX}.\
Such time-average statistical mechanics 
focuses on
functionals of a trajectory $(\x_\tau)_{0\le\tau\le t}$, in particular
the empirical density (or occupation time
\cite{Kac1949TAMS,Darling1957TAMS,Aghion2019PRL,Carmi2011PRE,Majumdar2002PRL,Majumdar2002PREa,Majumdar2005CS,Bray2013AP,Bel2005PRL}) 
$\overline{\rho_\x}(t)$ and current $\overline{\f J_\x}(t)$ at a point
$\x$.~Necessary in the analysis of laboratory \cite{Battle2016S,Gladrow2016PRL} or computer
\cite{Li2019NC} experiments with a finite spatial resolution, and
  useful for 
  smoothing 
  data \emph{a posteriori} to improve statistics,
the density and current 
should be defined as spatial
averages over a window $U^h_\x(\x')$ at $\x$ with coarse-graining scale $h$
\begin{align} 
\overline{\rho^U_\x}(t)&\equiv \frac{1}{t}\int_0^t
U^h_\x(\x_\tau)\rmd\tau\nonumber\\
\overline{\f J^U_\x}(t)&\equiv \frac{1}{t}\int_{\tau=0}^{\tau=t}U^h_\x(\x_\tau)\circ\rmd\x_\tau,
\label{def_current}
\end{align}
where $\circ\,\rmd\x_\tau$ denotes the Stratonovich integral. These
observables are illustrated in \blue{terms of sojourns of the window in Fig.~\ref{Fg1}a,b. 
Choosing the window $U^h_\x$ as a bin, the density and current
observables appear as histograms along single trajectories over
occupations of or displacements in the bin that  fluctuate between different realizations (see Fig.~\ref{Fg1}c-e and accompanying extended paper \footnote{See
accompanying extended paper available at  \href{https://arxiv.org/abs/2204.06553}{https://arxiv.org/abs/2204.06553}}).} Aside from coarse graining, the integration over 
  $U^h_\x(\x')$ may also represent a 
  pathwise thermodynamic
  potential\blue{, e.g.~heat dissipation (the force integrated along a
    stochastic path $\int_{\tau=0}^{\tau=t}\f F(\x_\tau)\cdot\circ\rmd\x_\tau$ \cite{Qian_PRE}) or generalized currents \cite{Dechant2018JSMTE,Dechant2021PRR,Dechant2021PRX,Dechant2018JPAMT}}.~Normalized windows, i.e.\ $\int {U^h_\x}(\z)d\z
 =1$, yield $\overline{\rho^U_\x}(t)$ and $\overline{\f J^U_\x}(t)$
 that are
 estimators of the probability density and current density,
 respectively.~The usually defined empirical density,
   $\overline{\rho_\x}(t),$ and current, $\overline{\f J_\x}(t)$,
 \cite{Maes2008PA,Touchette2009PR,Kusuoka2009PTRF,Chetrite2013PRL,Chetrite2014AHP,Barato2015JSP,Hoppenau2016NJP,Touchette2018PA,Mallmin2021JPAMT,Monthus2021JSMTE}
 correspond to no coarse graining, i.e.~$U^{h=0}_\x(\z)$ being 
 Dirac's delta function $\delta(\x-\z)$.\\
\emph{Reliably} inferring from noisy trajectories
whether a system obeys detailed balance, notwithstanding recent
progress
\cite{Roldan2010PRL,Fodor2016PRL,Battle2016S,Gladrow2016PRL,Gnesotto2018RPP,Li2019NC,Gladrow2019NC,Berezhkovskii2020JPCL,Seara2021NC},
remains challenging.~\emph{Quantifying} violations of detailed balance
is a daunting task.~One can quantify broken detailed
balance through violations of the fluctuation dissipation theorem
\cite{Cugliandolo1997PRL,Mizuno2007S,Seifert2010EEL}, which requires
perturbing the system from the steady state. 
One can also
check for a symmetry breaking of forward/backward transition-path
times \cite{Gladrow2019NC,Berezhkovskii2020JPCL}, measure
the entropy production
\cite{Roldan2010PRL,Seifert2012RPP,Pigolotti2017PRL,Seara2021NC}, or
infer steady-state currents (see arrows in Fig.~\ref{Fg1}c) directly
\cite{Battle2016S,Gladrow2016PRL}, all of which require substantial statistics.~However, single-molecule experiments often cannot reach ergodic
times, have a finite resolution,
and only allow for a limited number of repetitions.~This leads to
uncertainties in estimates of observables such as
steady-state currents (see Fig.~\ref{Fg1}d-f).~Notably,
  fluctuations of $\overline{\rho^U_\x}$ and
  $\overline{\f J_\x^U}$ encode information about violations of
    detailed balance (even where the \blue{mean} current or its components locally
    vanish;\ see Fig.~\ref{Fg1}d,e), which \emph{a priori} is hard
to interpret.\\
\indent Current fluctuations have a noise
floor---they are bounded from below by the ``thermodynamic
uncertainty relation''
\cite{Barato2015PRL,Gingrich2016PRL,Dechant2018JSMTE}
which
  in turn allows for bounding 
  dissipation in a system from below by current fluctuations
  \cite{Gingrich2017JPAMT,Seifert2018PA,Otsubo2020PRE,Supriya}.~As we
  show in Fig.~\ref{Fg1}f (see \cite{Note1} for a multi-well potential)
  the precision of inferring
dissipation \blue{typically} depends non-monotonically on the coarse-graining scale
$h$---given a system, a point $\x$, and trajectory length $t$ there
exists
a thermodynamically ``optimal'' coarse graining due to a diverging
variance for $h\to 0$ and vanishing mean for large $h$.~Moreover, $\overline{\rho_\x}$ and
$\overline{\f J_\x}$ without coarse graining turn out to be
ill-defined.\\
\indent In systems and on time scales where dynamics is reasonably
  described  by a Markov jump process on a small state space, 
  current fluctuations are well understood
  \cite{Lebowitz1999JSP,Qian2002PNASU,Pilgram2003PRL,Bodineau2004PRL,Andrieux2007JSP,Bertini2005PRL,Zia2007JSMTE,Maes2008EEL,Baiesi2009JSP,Barato2015JSP,Pietzonka2016PRE,Gingrich2016PRL,Gingrich2017PRL,Kapfer2017PRL,Macieszczak2018PRL,Kaiser2018JSP,Barato2018NJP,Marcantoni2020PRE}. 
  However, dynamics typically evolves in continuous
  space, and a continuous dynamics observed on a discrete space
  is \emph{not} Markovian \cite{Hartich2021PRX,Suarez2021JCTC} (see \footnote{See Supplemental Material at [...]
for \blue{further details and auxiliary results}, as well as Refs. \cite{SLapolla2021PRR,SSuarez2021JCTC,SGardiner1985,SHolubec2019PRE,SGingrich2017JPAMT,SymPy,ArxivJPhysA}.} for a quantitative confirmation).~An accurate Markov jump
  description may require too many states to
  be practical, and is known to fail when considering
  functionals as
  in Eq.~\eqref{def_current} \cite{Suarez2021JCTC}.\ 
  We therefore focus on continuous space, where, with 
  exceptions
  \cite{Flandoli2005SPA,Maes2008PA,Chernyak2009JSP,Dechant2018JSMTE},
  insight is 
  limited to hydrodynamic scales
  \cite{Bodineau2004PRL,Bertini2005PRL,Bertini2015RMP} and large
  deviations
  \cite{Touchette2009PR,Kusuoka2009PTRF,Chetrite2013PRL,Chetrite2014AHP,Barato2015JSP,Hoppenau2016NJP,Touchette2018PA,Mallmin2021JPAMT,Monthus2021JSMTE}. 
A comprehensive understanding of fluctuations and spatial correlations
of density and current in continuous space
remains elusive, and the interpretation 
\finalchange{of the typical definition without coarse graining in dimensions $d\ge 2$} apparently requires a revision, see below.

Here, we provide general results on
the empirical density and current in overdamped diffusive steady-state
systems, revealing a mathematical, thermodynamical, and
  experimental necessity for spatial coarse graining. When
defined in a point, fluctuations are proven to diverge in spatial 
dimensions above one, contradicting 
existing \finalchange{central-limit} statements.~We explain why a systematic
variation of the coarse-graining scale provides deeper insight about the
underlying dynamics and allows for improved inference of the
  system's thermodynamics.~Exploiting a \blue{generalized time-reversal symmetry} 
  we provide
intuition 
about fluctuating currents along
individual trajectories.~Non-vanishing density-current correlations 
are shown to unravel 
violations of
detailed balance from short
measurements.~Our results 
allow for a more consistent and
efficient analysis 
of 
experiments, and provide new insight into non-equilibrium steady states and their thermodynamics.

\blue{\emph{Setup.---}We consider time-homogeneous overdamped Langevin dynamics \cite{Gardiner1985,Pavliotis2014TiAM} in $d$-dimensional space evolving according to the stochastic differential equation
${\rmd\x_\tau=\f
F(\x_\tau)\rmd\tau+\bsig\rmd\f W_\tau}$, where
${\rmd\f W_\tau}$ is the increment of a $d$-dimensional Wiener
processes (i.e.~white noise) with covariance
$\langle\rmd W_{\tau,i}\rmd W_{\tau',j}\rangle=\delta(\tau-\tau')\delta_{ij}\rmd \tau\rmd
\tau'$.} The {Fokker-Planck} 
equation for the conditional probability density 
with initial
condition ${G(\x,0|\y)=\delta(\x-\y)}$ reads
${(\partial_t+\nabla_{\x}\cdot\bj_{\x})G(\x,t|\y)=0}$ with current
operator $\bj_\x\equiv \f F(\x)-\blue{\f D}\nabla_{\x}$, where \blue{$\f D\equiv\bsig\bsig^T/2$} is the 
positive definite diffusion matrix. \blue{All results directly 
  generalize to multiplicative noise (see \cite{Note1})}. The drift $\f F(\x)$ is assumed to be sufficiently smooth and
confining to ensure the
existence of a steady-state
density ${G(\x,t\!\to\!\infty|\y)\!=\!\ps(\x)}$ and, if detailed
balance is violated, a steady-state current
${\js(\x)\equiv\!\bj_\x\ps (\x)\ne \f 0}$ \cite{Gardiner1985,Pavliotis2014TiAM}.
\begin{figure*}[ht!!]
\begin{center}
\includegraphics[width=1.0\textwidth]{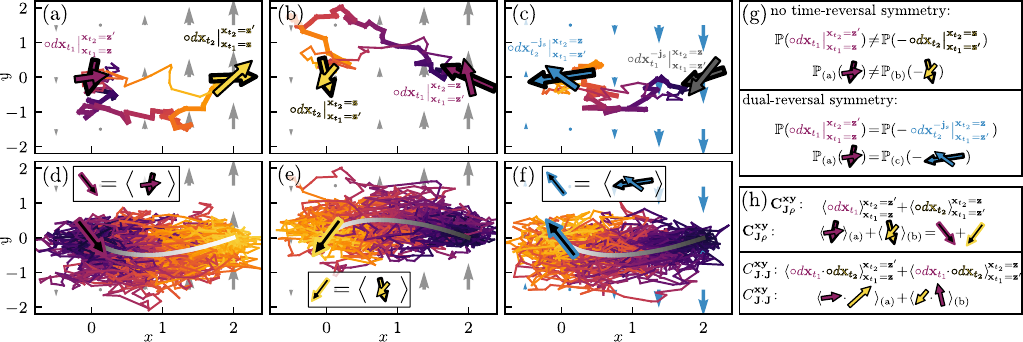}
\caption{(a)~Two sample trajectories in a shear flow $\f F_{\rm sh}(\x)$ (grey
    arrows) with Stratonovich displacements 
    $\circ\rmd\x_{t}$ in the
    initial $\x_{t_1}=\z$ and final point $\x_{t_2}=\z'$ for fixed
    $t_1<t_2$ depicted
    by purple and yellow arrows, respectively.~Time is running
    from dark to bright.~(b)~Trajectories as in (a)
    but running from $\x_{t_1}=\z'$ to $\x_{t_2}=\z$.~(c)~As in
    (b) but with the inverted shear flow $-\f F_{\rm sh}(\x')$
    (blue background arrows) and initial and final increments depicted
    by grey and blue arrows.~(d)~Ensemble of paths
    from $\x_{t_1}=\z$ to $\x_{t_2}=\z'$ contributing to 
    $P_\z(\z',t_2-t_1)$.~The average
    initial displacement
    $\E{\circ\rmd\x_t}_{\x_{t_1}=\z}^{\x_{t_2}=\z'}$
    is depicted by the black-purple arrow, and the mean path
    $\z\to\z'$ in time $t_2-t_1$ by the gray gradient line.~(e)~As in (d) but
    corresponding to (b) instead of (a).~(f)~As in (e) but with the
    reversed shear flow as in
    (c).~(g-h)~Since the shear flow breaks time-reversal symmetry, initial-point
    increments in (a) cannot be obtained by
    inverting
    final-point increments in (b).~By dual-reversal symmetry initial-point
    increments follow from inverting the final-point increments
    in the inverted shear flow in (c), which explains initial point
    increments $\circ\rmd\x_{t_1}$ in current-density correlations and current (co)variances
    via the easier and more intuitive final point increments $\circ\rmd\x^{-\js}_{t_2}$.
  \label{Fg2}}
\end{center}
\end{figure*}

\blue{\emph{Correlations and fluctuations from paths.---}To
  investigate the non-trivial statistics of the observables in
  Eq.~\eqref{def_current} we now outline the derivation detailed in
  \cite{Note1} of results for mean values, correlations and
  fluctuations assuming steady-state initial conditions.
  Let $\langle\cdot\rangle_{\rm s}$ denote the average over all paths
  $\{\x_\tau\}$ evolving from $\ps$.\ The mean values
  ${\langle\overline{\rho_\x^U}(t)\rangle_{\rm s}=\int d\z
    U^h_\x(\z)\ps(\z)}$ and ${\langle\overline{\f
      J^U_\x}(t)\rangle_{\rm s}=\int d\z U^h_\x(\z)\js(\z)}$
  \cite{Note1} are time-independent estimators of the steady-state
  density and current coarse-grained over a window $U^h_\x$.\ In
  contrast to the mean values, covariances display a non-trivial
  time-dependence and therefore contain salient features of the
  dynamics.} We define the two-point steady-state covariance as
\begin{equation}
C^{\x\y}_{AB}(t)\equiv\langle A_\x(t)B_\y(t)\rangle_{\rm s} - \langle
A_\x(t)\rangle_{\rm s}\langle B_\y(t)\rangle_{\rm s}\, ,
\label{covar_g}
\end{equation}
where $A$ and $B$ are either $\overline{\rho^U}$ or $\overline{\f
  J^U}$, respectively. We refer to the case $A\ne B$ or $\x\ne\y$ as
(linear) correlations and to $A=B$ with $\x=\y$ as fluctuations
with the \blue{notation} ${{\rm var}^\x_A(t)\equiv
  C^{\x\x}_{AA}(t)}$. \blue{Recall that ${\rm var}^\x_\rho(t)$ and
  ${\rm var}^\x_{\f J}(t)$ quantify (experimentally relevant)
  fluctuations of histograms along single trajectories (see
  Fig.~\ref{Fg1}d,e), and ${\rm var}^\x_{\f J}(t)$ is at the heart of the thermodynamic uncertainty relation (see Fig.~\ref{Fg1}f). Moreover, $C^{\x\y}_{\f J \rho}(t)$ was recently found to play a vital role in stochastic thermodynamics \cite{Dechant2021PRX}. All $C^{\x\y}_{AB}(t)$ are easily inferred from data, but lack physical understanding.}
    We now give $C^{\x\y}_{AB}(t)$ a physical meaning in terms of the statistics of
    paths pinned at end-points $\z$ and $\z'$ (see
    Fig.~\ref{Fg2}).~Introduce $\E{\cdot}^{\x_{t_2}=\z'}_{\x_{t_1}=\z}\equiv\E{\,\delta(\x_{t_1}-\z)\delta(\x_{t_2}-\z')\,\cdot}_{\rm
      s}$, the Stratonovich increment 
$\!{\circ\rmd\x_\tau\equiv\x_{\tau+\rmd\tau/2}\!-\!\x_{\tau-\rmd\tau/2}}$,
    and the 
    operator
\begin{align}
\integrals^{t,U}_{\x\y}[\cdot]\equiv \frac{1}{t^2}\!\int_0^tdt_1\int_{t_1}^tdt_2\int\!d\z U^h_\x(\z)\!\int\!d\z' U^h_\y(\z')[\cdot],
\label{int_op}
\end{align}
\blue{where $[\cdot]$ represents functions of $t_1,t_2,\z,\z'$ and without loss of generality we choose} the convention
$\int^t_{t_1}dt_2\delta(t_2-t_1)=1/2$.~\blue{Upon plugging in mean
  values $\Es{A_\x}$ and $\Es{B_\y}$,
  the
  definition~\eqref{covar_g}  becomes} \cite{Note1} $C^{\x\y}_{\rho\rho}(t)=\integrals^{t,U}_{\x\y}[\Xi^{\z\z'}_1-2\ps(\z)\ps(\z')]$
for density-density correlations,
$\f C^{\x\y}_{\f J   \rho}(t)=\integrals^{t,u}_{\x\y}[\boldsymbol{\Xi}^{\z\z'}_2-2\js(\z)\ps(\z')]$
for current-density correlations, and (see \footnote{Fluctuations and correlations
of $\overline{\f J^U_\x}(t)$ are characterized by the $d\times d$ covariance matrix with elements $(\mathsf{C}^{\x\y}_{\f J \f J}(t))_{ik}=C^{\x\y}_{{\rm J}_i  {\rm J}_k}(t)$. We focus on the scalar case $C^{\x\y}_{\f J\cdot \f J}(t)\equiv \mathrm{Tr}\mathsf{C}^{\x\y}_{\f J \f J}(t)$. See \cite{ArxivJPhysA} for the full covariance matrix.})
$C^{\x\y}_{\f J\cdot \f
  J}(t)=
\integrals_{\x\y}^{t,U}[\Xi^{\z\z'}_3-2\js(\z)\cdot\js(\z')]$
for
current-current correlations,
where we defined
\begin{eqnarray}  
  &&\blue{\Xi^{\z\z'}_1\equiv
   {\E{1}}^{\x_{t_2}=\z'}_{\x_{t_1}=\z}
 +
{\E{1}}^{\x_{t_2}=\z}_{\x_{t_1}=\z'}}
\nonumber\\
&&\boldsymbol{\Xi}^{\z\z'}_2\equiv
  \frac{{\E{\circ\rmd\x_{t_1}}}^{\x_{t_2}=\z'}_{\x_{t_1}=\z}}{\rmd
    t_1}+\frac{{\E{\circ\rmd\x_{t_2}}}^{\x_{t_2}=\z}_{\x_{t_1}=\z'}}{\rmd
    t_2}
  \label{ansatz_increments}\\
&&\Xi^{\z\z'}_3\equiv\frac{{\E{\circ\rmd\x_{t_1}\cdot\circ\rmd\x_{t_2}}}^{\x_{t_2}=\z'}_{\x_{t_1}=\z}}{\rmd
    t_1\rmd t_2}+\frac{{\E{\circ\rmd\x_{t_1}\cdot\circ\rmd\x_{t_2}}}^{\x_{t_2}=\z}_{\x_{t_1}=\z'}}{\rmd
    t_1\rmd t_2}\,.\nonumber
\end{eqnarray}
\blue{Eqs.~\eqref{int_op}-\eqref{ansatz_increments} tie
  $C_{AB}^{\x\y}$ to properties of pinned paths, weighted by
  $U_\x^h(\z),U_\y^h(\z')$ and integrated over space and times $0\le
  t_1\le t_2\le t$. In contrast to the somewhat better
  understood density-density covariance
  \cite{Lapolla2020PRR,Lapolla2018NJP,Kac1949TAMS}, current-density
  and current-current covariances involve (scalar products of) more
subtle Stratonovich increments along pinned trajectories, explained graphically in Fig.~\ref{Fg2} and further investigated in the following.

\emph{Correlations and fluctuations from two-point densities.---}To obtain quantitative results, we evaluate the averages ${\E{\cdot}}^{\x_{t_2}=\z'}_{\x_{t_1}=\z}$ in terms of two-point functions $P_{\z}(\z',t_2-t_1)\equiv G(\z',t_2-t_1|\z)\ps(\z)$.\ For density-density correlations $C^{\x\y}_{\rho\rho}$ the result is readily obtained from Eq.~\eqref{ansatz_increments} using ${\E{1}}^{\x_{t_2}=\z'}_{\x_{t_1}=\z}=P_{\z}(\z',t_2-t_1)$}.
Conversely,
Stratonovich increments, are difficult to understand and hard to
evaluate, particularly \emph{initial-point
increments $\circ\rmd \x_{t_1}$ because they are correlated with 
future events.}\ \\
\indent To gain intuition we examine a two-dimensional shear flow
$\f F_{\rm sh}(\x)=2x\hat{\y}$ 
shown
in Fig.~\ref{Fg2}, depicting initial-, $\circ\rmd \x_{t_1}$, and
end-point, $\circ\rmd \x_{t_2}$, increments along forward (Fig.~\ref{Fg2}a) and time-reversed
(Fig.~\ref{Fg2}b) pinned trajectories between times $t_1<t_2$ 
and their
ensemble averages
(Fig.~\ref{Fg2}d-e).\ In the accompanying extended paper \cite{Note1}
  we show that $\E{\circ \rmd
    \x_{t_2}}^{\x_{t_2}=\z'}_{\x_{t_1}=\z}=\bj_{\z'}
  P_{\z}(\z',t_2)\rmd t_2$\blue{, i.e.\  mean displacements are given
    by the Fokker-Planck current as expected}.\ Moreover, when
  detailed balance holds, \blue{time-reversal symmetry implies} ${\mathbbm{P}(\circ\rmd
\x_{t_1}|^{\x_{t_2}=\z'}_{\x_{t_1}=\z})=\mathbbm{P}(-\circ\rmd \x_{t_2}|^{\x_{t_2}=\z}_{\x_{t_1}=\z'})}$, whereas under broken
detailed balance, e.g.~due to the shear flow in
Fig.~\ref{Fg2}, this ceases to hold.
We may, however, employ a generalized time-reversal symmetry---the
\emph{dual-reversal symmetry} (see~\cite{Note1} and
\cite{Sasa2014JSMTE,Dechant2021PRR,Hatano2001PRL})---implying ${\mathbbm{P}(\circ\rmd
\x_{t_1}|^{\x_{t_2}=\z'}_{\x_{t_1}=\z})=\mathbbm{P}(-\circ\rmd
\x^{-\js}_{t_2}|^{\x_{t_2}=\z}_{\x_{t_1}=\z'})}$ connecting ensembles 
with currents
$\js$ and $-\js$ (see Fig.~\ref{Fg2}c,f,g).~\blue{Via this generalized time-reversal symmetry} we circumvent the correlation of
$\circ\rmd \x_{t_1}$ with the
future.\ To materialize this we 
isolate the
irreversible drift in $\bj_\x=\ps(\x)^{-1}\js(\x)-\ps(\x)\blue{\f
D}\nabla_\x\ps(\x)^{-1}$, and introduce the dual current operator
 $\revj_{\x}\equiv-\bj^{-\js}_{\x}=\ps(\x)^{-1}\js(\x)+\ps(\x)\blue{\f
D}\nabla_\x\ps(\x)^{-1}$,
rendering all terms in Eq.~\eqref{ansatz_increments} (illustrated in
Fig.~\ref{Fg2}h) tractable,  and ultimately leading to our main result
\begin{eqnarray}
&&\f C^{\x\y}_{\f J \rho}(t)=\integrals^{t,U}_{\x\y}[\bj_\z
  P_{\z'}(\z,t')+\revj_\z
  P_{\z}(\z',t')-2\js(\z)\ps(\z')]\nonumber\\
&&C^{\x\y}_{\f J\cdot \f J}(t)=\blue{\frac{2{\rm Tr}\f D}{t}\int d\z}U^h_\x(\z)U^h_\y(\z)\ps(\z)+\label{var_j_j}\\
&&\integrals^{t,U}_{\x\y}[\bj_\z\cdot\revj_{\z'}P_{\z'}(\z,t')+\bj_{\z'}\cdot\revj_\z
  P_{\z}(\z',t')-2\js(\z)\cdot\js(\z')],\nonumber
\end{eqnarray}
where the first term in $C^{\x\y}_{\f J\cdot \f J}(t)$ arises from \blue{$t_1=t_2$~\cite{Note1},
and the operator
$\integrals^{t,U}_{\x\y}$ simplifies $t^{-2}\int_0^tdt_1\int_{t_1}^tdt_2\to t^{-1}\int_0^t dt'(1-t'/t)$ since Eq.~\eqref{var_j_j} depends only on time differences $t'\equiv t_2-t_1\ge0$. Notably, written in this simplified form Eq.~\eqref{var_j_j} establishes Green-Kubo relations \cite{Green,Kubo} connecting covariances $C_{AB}^{\x\y}$ to time-integrals of generalized correlation functions.}

\begin{figure}[ht!]
\includegraphics[width=0.47\textwidth]{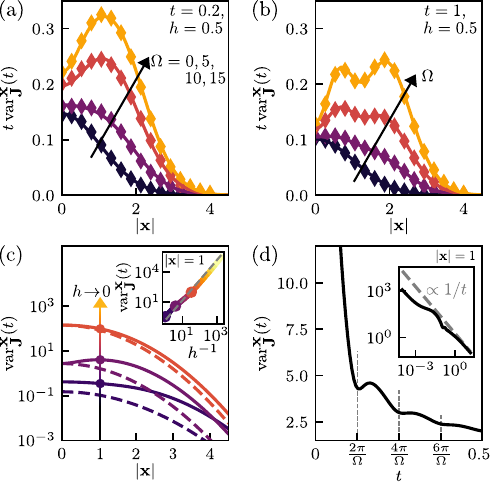}
\caption{$t\,{\rm var}_{\f J}^\x$ as a function of the radius
  $|\x|$ in the harmonically confined rotational flow in Fig.~\ref{Fg1}c for
  increasing
  $\Omega$ with Gaussian $U^h_\x$ with width $h$ at (a)
  $t=0.2$ and (b) $t=1$;~Lines depict Eq.~\eqref{var_j_j} and
  symbols
   simulations~\cite{Note2}.~(c) $t\,{\rm var}_{\f J}^\x$
  at $t=1$ for $\Omega=10$ (full lines) and equilibrium $\Omega=0$
  (dashed lines), for various $h$ decreasing along the
  arrow.~Inset:~divergence of ${\rm var}_{\f J}^\x$ as $h\to 0$ at
  $|\x|=1$;~the dashed line depicts Eq.~\eqref{bound local time}.~Note
  the logarithmic scales.~(d) ${\rm var}_{\f J}^\x$ as a function
  of $t$ for very strong driving $\Omega=50$; Inset: (d) on logarithmic
  scales alongside the \finalchange{central-limit} scaling $\propto t^{-1}$.}
\label{Fg3}
\end{figure}

\blue{
Given the two-point function $P_\z(\z',t')$, Eq.~\eqref{var_j_j} gives
the correlation and fluctuations of observables defined in
Eq.~\eqref{def_current}.\ In practice, $P_\z(\z',t')$ may not
necessarily  be
available.\ However, the theoretical result Eq.~\eqref{var_j_j}
nevertheless allows us to draw several conclusions, in particular by
considering special cases and limits.}\
At equilibrium 
$\revj_\z=-\bj_\z$, 
implying $\f C^{\x\y}_{\f J \rho}(t)=0$.\ A non-zero $\f C^{\x\y}_{\f J
  \rho}(t)$ at any time $t$ 
  is thus
a conclusive signature of broken detailed balance. Moreover, at equilibrium  $C^{\x\y}_{\f J\cdot \f J}(t)$
does not vanish 
although $\langle\overline{\f J_\x^U}\rangle_{\rm s}=0$.
When $\js\ne \f 0$, ${\rm var}_{\f J}^\x(t)\equiv C^{\x\x}_{\f J\cdot \f J}(t)$ may
display maxima where $P_{\rm s}(\x)$ has none (see
Fig.~\ref{Fg3}a-c), and an oscillatory time-dependence due to
circulating currents 
(see Fig.~\ref{Fg3}d), both signaling
non-equilibrium.
\blue{For a more detailed discussion of Eq.~\eqref{var_j_j} see \cite{Note1}.}

\blue{\emph{Necessity of coarse graining.---}Of particular interest is the dependence of fluctuations on the coarse-graining length scale $h$ (see Fig.~\ref{Fg1}f, Fig.~\ref{Fg3}c and \cite{Note1}). Importantly, the limits $h\to\infty$ and $h\to 0$ are generally accessible from Eq.~\eqref{var_j_j} independent of the detailed dynamics (see \cite{Note1}).
The limit $h\to 0$ with $U^h_\x(\z)\to\delta(\x-\z)$ corresponds to no coarse graining, i.e.\ the observables Eq.~\eqref{def_current} are evaluated in a single point $\z$. In this limit, the variance and covariance of
$\overline{\rho^U_\x}$ and $\overline{\f J^U_\x}$ for $d\ge2$
  and any $t$ behave as \cite{Note1}
\begin{align}
\text{var}^\x_\rho(t)\!&\overset{h\to 0}{\simeq}\blue{\frac{k\ps(\x)}{t}\times}
\begin{cases} \frac{h^{2 - d}}{d - 2} & \text{for}\: d > 2 \\
  -\ln{h} & \text{for}\: d = 2 \end{cases}
\nonumber\\
\f C^{\x\x}_{\f J\rho}(t)\!&\overset{h\to 0}{\simeq}
\js(\x)\text{var}^\x_\rho(t)/2\ps(\x)\nonumber\\
\text{var}^{\x}_{\f J}(t)\!&\overset{h\to 0}{=}\!\blue{
\frac{k'\ps(\x)}{t}}(d-1)h^{-d}\!+\mathcal{O}(t^{-1})\mathcal{O}(h^{1-d}),\label{bound local time}
\end{align}
where $\simeq$ denotes asymptotic equality\blue{, and $k,k'$ are constants depending on $\f D$ and $U_\x$ \cite{Note1}. Therefore, 
taking $U^{h}_\x(\z)\overset{h\to 0}{\longrightarrow}\delta(\x-\z)$ as
implicitly assumed in
\cite{Touchette2009PR,Kusuoka2009PTRF,Chetrite2013PRL,Chetrite2014AHP,Barato2015JSP,Hoppenau2016NJP,Touchette2018PA,Lapolla2020PRR,Mallmin2021JPAMT,Monthus2021JSMTE}
we find for $d\ge 2$ that $\text{var}^{\x}_{\rho,\f J}(t),\f
C^{\x\x}_{\f J\rho}(t)$ diverge for all $t$ (see 
Fig.~\ref{Fg3}c). Eq.~\eqref{bound local time} also applies to
Markov-jump processes defined on a grid with spacing $h\to 0$; for
details and an example see \cite{Note2}.}
The divergence can be understood intuitively \cite{Note1}, e.g.\ based on the following
argument.\\
\indent Note that the probability that point $\z$ is hit by the trajectory
$(\x_\tau)_{0\le\tau\le t}$, i.e.\ that there is a $\tau\in[0,t]$ such
that $\x_\tau=\z$, delicately depends on the spatial dimensionality
$d$. This probability is positive for $d=1$ but zero in
higher-dimensional space. That is,  $\mathbbm{P}({\exists
\tau\in(0,t]\colon\x_\tau=\z})=0$ for diffusion in $d\ge2$
\cite{Note1,Durrett_Stoch}.\ Mean values remain finite in the limit $h\to 0$, namely $\langle\overline{\rho_\x}(t)\rangle_{\rm s}=\ps(\x)$ and $\langle \overline{\f
J_\x}(t)\rangle_{\rm s}=\js(\x)$ in agreement with existing literature
\cite{Maes2008PA,Chernyak2009JSP,Chetrite2014AHP,Barato2015JSP,Hoppenau2016NJP,Touchette2018PA,Mallmin2021JPAMT,Monthus2021JSMTE}. Since
the probability to hit the point $\z$ is approaching zero as
$h\to 0$, this implies that the mean is precisely balanced by the
infinite contribution of the delta  
function $U^h_\x(\z)\to\delta(\x-\z)$, as in
$\Es{\delta(\x_\tau-\x)}=\ps(\x)$.\ Loosely speaking, here
``$0\times\infty$'' is finite.  One may therefore expect diverging
second (and higher) moments when $h\to 0$ as this argument 
extends to ``$0\times\infty^2=\infty$''.\ The argument is not limited
to overdamped motion but seems to extend to a larger class of
stochastic dynamics, such as underdamped diffusion and experimental
data on anomalous intracellular transport \cite{Sabri2020PRL} shown in
Fig.~9 of Ref.~\cite{Note1}.} 

\blue{We hypothesize that not only the moments diverge, but that the
  density and current cannot even be consistently defined for $h=0$. Moreover, the 
limits $h\to 0$ and $t\to\infty$ do \emph{not} commute.\ This has
important consequences for \finalchange{the central-limit regime}, i.e.~statistics on
longest time scales (see Appendix~II and \cite{Note1}).\ Some coarse
graining $h>0$ is therefore necessary for mathematical consistency 
and anticipated \finalchange{central-limit} properties.}

Notably, for small windows Eq.~(\ref{bound local
  time}) implies that fluctuations (unlike correlations)
carry no information about steady-state currents $\js(\x)$ and
  thus 
  violations of detailed balance and
  thermodynamic properties such as the system's dissipation.\ In this
limit fluctuations reflect only Brownian, thermal currents 
that are invariant with respect to $\js(\x)$---systems
with equal $\ps (\x)$ and $\blue{\f D}$ display identical 
fluctuations (see~Eq.~\eqref{bound local
  time} and Fig.~\ref{Fg3}c).\ \blue{Recall that the dissipation can be inferred from current fluctuations via the thermodynamic uncertainty relation \cite{Barato2015PRL,Gingrich2016PRL,Gingrich2017JPAMT}.\ We now see that
 only an intermediate coarse
  graining,
  such as the ``optimum'' in Fig.~\ref{Fg1}f, allows
  to infer dissipation from fluctuations.
  Moreover, spatial features of steady-state currents
  (see~Fig.~\ref{Fg3}c) are only revealed with coarse graining. Some
  coarse graining $h>0$  is thus
  necessary to infer
  thermodynamic properties. In addition, divergent fluctuations make it
  impossible to accurately infer densities and currents without coarse
  graining from
  experiments.\ Experiments also nominally have a finite spatial
  resolution.\ Thus, coarse graining is also experimentally necessary.

\emph{Conclusion.---}}Leveraging It\^o calculus and
generalized time-reversal symmetry we were able to provide elusive
physical intuition about fluctuations and correlations of empirical
densities and currents that are
central to stochastic thermodynamics.
We established the so far overlooked necessity for spatial
coarse graining---it is required to ensure
mathematically well defined observables and the validity of \finalchange{central-limit statements} in dimensions $d\ge 2$, to improve the accuracy of
inferring thermodynamic properties (e.g.~dissipation) from
fluctuations and to uncover salient features of non-equilibrium steady-state currents
without inferring these individually
\cite{Frishman2020PRX,Brueckner2020PRL,Ferretti2020PRX}, and is
unavoidable in the analysis of experimental data with a finite resolution.\ Non-vanishing current-density correlations were
shown to be a conclusive indicator of broken detailed balance, and may
improve the accuracy of inferring invariant densities
\cite{Hartich2016PRE} and dissipation far from
equilibrium \cite{Dechant2021PRX}.
Our results allow for generalizations to
non-stationary initial conditions or non-ergodic dynamics, which will be addressed in forthcoming publications. 

\emph{Acknowledgments.---}Financial support from Studienstiftung des Deutschen Volkes (to C.\ D.) and the German Research Foundation (DFG) through the Emmy Noether Program GO 2762/1-2 (to A.\ G.) is gratefully acknowledged.

\blue{\emph{Appendix~I:~Density and current from sojourns.---}In
  general the density and current functionals measure the ($U^h_\x$-weighted) 
time spent and displacement accumulated in the window $U^h_\x$
averaged over time. Specifically, when $U^h_\x$  is the indicator
function, $U^h_\x(\z)=h^{-d}\mathbbm{1}_{\Omega_{\x}}(\z)$, of a region
$\Omega_\x$ centered at $\x$ with volume $h^d$, we can write this
illustratively in terms of the sojourns of the window as follows.
Letting the times of entering and exiting said window be $\tau_i^-$ and
 $\tau_i^+$, respectively, $ t\overline{\rho_\x^U}(t)$ corresponds to the sum of sojourn times, $\tau_i^s=\tau_i^+-\tau_i^-$, and $t\overline{\f J^U_\x}(t)$ the sum of vectors $\delta\x_i^s$ between entrance $\x_{\tau_i^-}$ and exit $\x_{\tau_i^+}$ points, that is, 
\begin{eqnarray}
  t\overline{\rho^U_\x}(t)&=&\frac{1}{h^d}\sum_{i\le N_t}(\tau_i^+-\tau_i^-)\equiv\frac{1}{h^d}\sum_{i\le N_t}\tau_i^s
\nonumber\\
  t\,\overline{\f J^U_\x}(t)&=&\frac{1}{h^d}\sum_{i\le N_t}(\x_{\tau_i^+}-\x_{\tau_i^-})\equiv\frac{1}{h^d}\sum_{i\le N_t}\delta\x^s_i,
  \label{current as sum of crossings}
\end{eqnarray}
where $N_t$ is the number of visits of the window. Note that
$N_t$ is almost surely either $\infty$ or 0, but the sum
converges.\ The points $\x_0$ or $\x_t$ 
may lie within $U^h_\x$ for which we set $\x_{\tau_{1}^-}=\x_0$ and/or
$\x_{\tau^+_{N_t}}\equiv\x_t$. As a result of correlations between
$\x_{\tau_{i}^-}$ and $\tau_i^s$ as well as $\x_{\tau_i^+}$ and
$\x_{\tau_{i+1}^-}$, $t\overline{\rho^U}$ and $t\,\overline{\f J^U}$
are in general \emph{not} renewal processes. A realization of
$\x_\tau$ in Fig.~\ref{Fg1}a,b provides intuition about
Eq.~\eqref{current as sum of crossings}.}

\emph{Appendix~II:~\finalchange{Central-limit regime}.---}\finalchange{Since the observables defined in Eq.~\eqref{def_current}
  involve time-averages, their statistics on the longest time scales
  is expected to be governed by the central limit theorem. Indeed, for
  non-zero $h$ or in spatial dimension $d=1$ (in both cases we
  obtained finite variances) on time scales $t$ that are very large
  compared to all time scales in the system, different parts of a
  trajectory (e.g.\ the sojourns in Fig.~\ref{Fg1}a and
  Eq.~\eqref{current as sum of crossings}) become sufficiently
  uncorrelated such that the central limit theorem implies Gaussian
  statistics. However, the diverging variance for $h\to 0$ for $d\ge
  2$ prevents Gaussian central-limit statistics on all time scales for
  the empirical density and current defined with a delta-function
  (i.e.\ without coarse graining). Since the diverging part of the variance in Eq.~\eqref{bound local time} has the dominant central-limit scaling $\propto t^{-1}$, the asymptotic variance $\sigma^{2}_{A}\overset{t\to\infty}{=}t\text{var}^\x_A(t)$ (where $A_\x(t)$ denotes $\overline{\rho^U_\x}(t)$ or $\overline{\f J^U_\x}(t)$) also diverges as $h\to 0$. This implies that taking $t\to\infty$ first and then $h\to 0$ also does \emph{not} yield finite variances. Moreover, note that the longest time scale in the system becomes the recurrence time, which diverges as $h\to0$. We hypothesize that a limiting distribution of $A_\x(t)$ only
exists as a scaling limit where $h\to 0$ and $t\to\infty$
simultaneously in some $d$-dependent manner \cite{Note1}. 

The central-limit regime is generally contained in the framework of large deviation theory \cite{Dembo2010,Touchette2009PR,Touchette2018PA}. Due to the divergent variance $\sigma_A^2$ and the resulting breakdown of Gaussian central-limit statistics, any large deviation principle for empirical densities and currents without coarse graining that predicts finite variances ceases to hold in $d\ge 2$.}

\let\oldaddcontentsline\addcontentsline
\renewcommand{\addcontentsline}[3]{}
\bibliographystyle{apsrev4-2}
\bibliography{bibPRLnew.bib}
\let\addcontentsline\oldaddcontentsline

\clearpage
\newpage
\onecolumngrid
\renewcommand{\thefigure}{S\arabic{figure}}
\renewcommand{\theequation}{S\arabic{equation}}
\setcounter{equation}{0}
\setcounter{figure}{0}
\setcounter{page}{1}
\setcounter{section}{0}

\begin{center}
\textbf{Supplementary Material for:\\{Mathematical, Thermodynamical, and Experimental Necessity
    for Coarse Graining Empirical Densities and Currents
  in Continuous Space}}\\[0.2cm]
Cai Dieball and Alja\v{z} Godec\\
\emph{Mathematical bioPhysics Group, Max Planck Institute for
  Multidisciplinary Sciences, Am Fa\ss berg 11, 37077 G\"ottingen}
\\[0.6cm]  
\end{center}

\begin{quotation}
\blue{In this Supplementary Material (SM) we present an explicit
  numerical confirmation that a discretely observed diffusion is
  \emph{not} a Markov process.Moreover, we apply the results for the limit of small window sizes to
  current fluctuations in a Markov jump process in discretized space.
Finally, we list the information that is necessary to reproduce all
simulations and analytical results shown in the figures presented the
Letter. Further details and derivations related to statements in the
Letter can be found in accompanying extended manuscript \cite{SAccompanyingPaper}.}
\end{quotation}
\hypersetup{allcolors=black}
\tableofcontents
\hypersetup{allcolors=mylinkcolor}

\section{Quantification of non-Markovianity of diffusion observed in discrete space (i.e.\ on a grid)}
In this section we provide a quantitative example for the statement
made in the Letter  that "a continuous dynamics observed on a discrete
space is \emph{not} Markovian". This in particular demonstrates that a
Markov-jump description with a limited number of states in general \emph{cannot}
accurately describe a diffusion process in continuous
space. A Markov-jump description may be accurate in systems with a
time-scale separation (e.g.\ as a result of high energy barriers
separating minima) on time scales sufficiently larger than the slowest
relaxation (e.g.\ much larger than the longest relaxation time in the
minima). However, in general an accurate Markov-jump representation
requires too many states to describe diffusive dynamics, and in particular
to accurately describe functionals of paths \cite{SSuarez2021JCTC}. The example we provide here is the
Ornstein-Uhlenbeck process with a rotational flow, see Fig.~1c in the
Letter. To quantify the non-Markovianity we discretize the dynamics on
a finite grid and quantify violations of the Chapman-Kolmogorov
equation, as described in Ref.~\cite{SLapolla2021PRR}.

\begin{figure}
\includegraphics[scale=1]{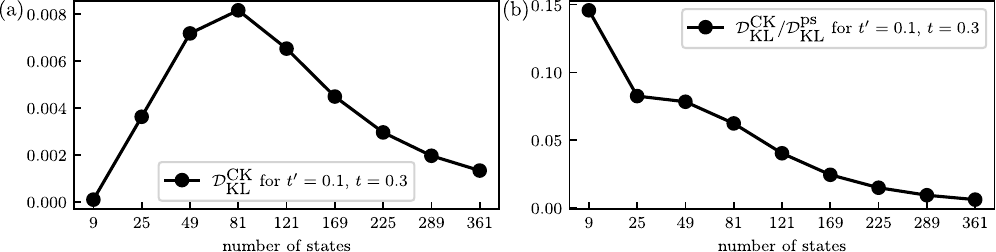}
\caption{Quantification of non-Markovianity (as described in the
    text) in a discrete-space observation with $N\times N$ states for
    $N=3,5,7,\dots,19$. The integrals in Eq.~\eqref{integrals
      rasterization} were evaluated numerically. (a) The value of
    $\mathcal D^{\rm CK}_{\rm KL}(t',t,i_0)$, for $i_0$ denoting the
    index of the square containing the point $(0,0)$, obviously
    decreases for large numbers of states according to
    expectations. For very small numbers of states, where very few
    details of the dynamics are observed, the
    non-Markovianity also appears to be smaller. (b) To gain intuition
    about the values of $\mathcal D^{\rm CK}_{\rm KL}(t',t,i_0)$ we
    normalized it by $\mathcal D^{\rm ps}_{\rm KL}(t,i_0)$, i.e.\ a
    value of $1$ means that the Kullback-Leibler divergence of $G$ and
    $G^{\rm CK}$ equals the divergence of $G$ and $G^{\rm ps}$. Since
    the Kullback-Leibler divergence is \emph{not} a metric (it is not
    symmetric and does not satisfy the triangle inequality), one
    should  be careful when interpreting its values \emph{quantitatively}.
\label{FgCK}}
\end{figure}

We consider the Ornstein-Uhlenbeck process in two-dimensional continuous space, i.e.\ Eq.~\eqref{OUP} with $r=D=1$ and $\Omega=3$. Then we divide the area $[-5,5]\times[-5,5]$ into $N\times N$ squares $S_i$ that we label by an index $i\in\{1,\dots N^2\}$. We define the steady-state occupation of the state $i$ (i.e.\ of the area/square $S_i$) from the continuous-space steady-state density $\ps(\x_0)$ (see Eq.~\eqref{ps and js}) as
\begin{align}
\ps(i_0)\equiv\int_{x_0\in S_{i_0}}\rmd^2x_0\ps(\x_0),
\end{align}
and, for any time $t$, from the continuous-space propagator (i.e.\ conditional density) $G(\x,t|\x_0)$ (see Eq.~\eqref{NESS-2d-OUP-propagator}) we define the propagator of the discrete space observation as
\begin{align}
G(i,t|i_0)\equiv\frac{1}{\ps(i_0)}\int_{x\in S_i}\rmd^2x\int_{x_0\in S_{i_0}}\rmd^2x_0 G(\x,t|\x_0)\ps(\x_0).\label{integrals rasterization} 
\end{align}
Moreover, we define \cite{SLapolla2021PRR}
\begin{align}
G_{t'}^{\rm CK}(i,t|i_0)\equiv\sum_{j=1}^{N^2}G(i,t-{t'}|j)G(j,{t'}|i_0).
\end{align}
Note that by the Chapman-Kolmogorov equation a Markov process would
obey $G_{t'}^{\rm CK}(i,t|i_0)=G(i,t|i_0)$ for all $i,i_0,t,t'$. To quantify non-Markovianity we compute the Kullback-Leibler divergence between $G$ and $G^{\rm CK}$,
\begin{align}
\mathcal D^{\rm CK}_{\rm KL}(t',t,i_0)\equiv\sum_{i=1}^{N^2}G(i,t|i_0)\ln\left[\frac{G(i,t|i_0)}{G_{t'}^{\rm CK}(i,t|i_0)}\right].
\end{align}
The results for this example are shown in Fig.~\ref{FgCK}a. Whenever
$\mathcal D^{\rm CK}_{\rm KL}> 0$ the process is
non-Markovian. However, the exact value of $\mathcal D^{\rm CK}_{\rm
  KL}\neq 0$ does not have a direct interpretation. To gain some
intuition about the actual value, in Fig.~\ref{FgCK}b we normalize by the Kullback-Leibler divergence of $G$ and $\ps$,
\begin{align}
\mathcal D^{\rm ps}_{\rm KL}(t,i_0)\equiv\sum_{i=1}^{N^2}G(i,t|i_0)\ln\left[\frac{G(i,t|i_0)}{\ps(i)}\right].
\end{align}
The rationale is that the actual value (at least) should not directly depend on how far the actual dynamics is displaced from the steady state, i.e.\  if the dynamics is closer to the steady state the same value of $\mathcal D^{\rm CK}_{\rm KL}$ should be interpreted as stronger violation of Markovianity as compared to when it is farther away, since the actual dynamics changes less in magnitude in the former case.

As expected, the extent of the violation of Markovianity depends
on the grid-size. It reduces for sufficiently small grids (large $N$), i.e.\ the
dynamics become effectively Markovian on shorter time-scales, as well
as for large ``ignorant'' discrete observations (small $N$), where all probability
flows are averaged over. Both limits are intuitive---non-Markovianity arises because there is no time-scale separation ensuring local-equilibrium. That is, the direction and rate of leaving a discretely observed state depends on the previous state
and the precise location of entering the binned discrete state. Notably (and obviously), the dynamics in the limit of infinite
number of grid points, i.e.\ in the continuum limit, is exactly the diffusion process and thus Markovian. Moreover, the attenuation of non-Markovianity for very
``ignorant'' grids (i.e.\ small number of states) as a result of
spatial averaging over large regions is easiest
understood by realizing that the dynamics with one state is (trivially) Markovian
and stationary at all times, the dynamics with two states slightly less so, etc.

\textbf{A diffusion observed on a grid is thus \emph{not} a Markov jump
process} and one in general requires many states for an
accurate discrete-state Markov-jump representation, which is typically not experimentally feasible. Moreover, one actually needs to parameterize
the Markov state model with such a large number of states, which is even
less feasible. \textbf{In contrast, evaluating empirical densities and
currents in finite windows assuming an underlying continuous-space
diffusion---as carried out in the present work---is \emph{not} constrained to small windows \emph{nor} does
it require any parameterization of a discrete-state model for its interpretation.}

\section{Implications for fluctuations in Markov jump processes on a grid}
If one approximates $d\ge2$-dimensional continuous-space
  dynamics by a Markov jump process on a grid, the fact that the
  Markov-jump description becomes asymptotically accurate for the
  number of states $N\to\infty$ implies that for large $N$, correlations and fluctuations of densities and currents will be governed by the limits for $h\to 0$ in Eq.~(6) in the Letter. \blue{Corresponding to the problems with large deviation theory for $h\to 0$ discussed in the appendix, central-limit large deviations in dimensions $d\ge 2$ exist only for finite grids (corresponding to $h>0$). In contrast, in the continuum limit where the number of states tends to infinity (see e.g.~\cite{SGingrich2017JPAMT}; corresponding to $h\to 0$) the recurrence time to visit a state diverges, which would require $t>\infty$ for validity of the large deviation principle. In particular, a finite relaxation time scale (where the relaxation time is the inverse of the smallest non-zero eigenvalue of the generator of the dynamics) that remains finite in the continuum limit alone does not guarantee the validity of the large deviation principle.

We now give a specific example for the validity of the limits for $h\to 0$ in Eq.~(6) in the Letter. }A state at position $\x$ of a Markov jump process on a grid with spacing $h$ can asymptotically be interpreted as corresponding to a window function $U^h_\x$ that is the normalized indicator function of a square with spacing $h$ around $\x$, e.g.\ in two-dimensional space $U^h_\x(x',y')=h^{-2}\mathbbm 1_{\abs{x-x'},\abs{y-y'}\le h/2}$. With this correspondence, applied to a two-dimensional example with constant isotropic diffusion, the current fluctuations for a small grid-spacing $h$ should \blue{according to Eq.~(6) in the Letter (see accompanying paper for prefactor $2D$) }be governed by
\begin{align}
{\rm var^\x_{\f J}}(t)\overset{h\to 0}=\frac{2D}{t}\ps(\x)(d-1)h^{-2}+\mathcal O(t^{-1})\mathcal O(h^{-1}).\label{limit grid} 
\end{align}
Rates $k_{i\to i'}^{j\to j'}$ (denoting $x$-indices by $i$ and $y$-indices by $j$) for a discretized process originating from continuous dynamics are not unique, but can e.g.\ be obtained following \cite{SHolubec2019PRE}. Via the pseudo potentials (we need to use \emph{pseudo} potentials since we consider a non-equilibrium process \cite{SHolubec2019PRE})
\begin{align}
\tilde U(x,y)&=\frac{r}{2D}x^2-\frac{\Omega}{D}xy\nonumber\\
\tilde V(x,y)&=\frac{r}{2D}y^2+\frac{\Omega}{D}xy,
\end{align}
we obtain rates
\begin{align}
k_{i\to i\pm 1}^{j\to j}&=\frac{D}{h^2}\exp\left(-\frac{1}{2}[\tilde U(x_{i\pm 1},y_j)-\tilde U(x_i,y_j)]\right)\nonumber\\
k_{i\to i}^{j\to j\pm 1}&=\frac{D}{h^2}\exp\left(-\frac{1}{2}[\tilde V(x_i,y_{j\pm 1})-\tilde V(x_i,y_j)]\right).
\end{align}
Note that all other rates (i.e.\ to non-neighboring states) vanish in this construction.

Currents are now defined as transition counts on the edges of the grid. To compare to continuous-space time-averaged currents, we define the $x$-component of a current $\overline{\f J_\x}$ at a grid point $\x$ as the net number of transitions on the edge to the right of $\x$ and the $y$-component as the net number of transitions on the edge above $\x$. Note that this current reflects probabilities of transitions and not probability densities which differs by the current density by a factor of $h^2$, i.e.\ by the area corresponding to of a state. With this definition, the limit Eq.~\eqref{limit grid} for $h\to 0$ becomes
\begin{align}
{\rm var^\x_{\f J}}(t)\overset{h\to 0}=\frac{2D}{t}\ps(\x)(d-1)+\mathcal O(t^{-1})\mathcal O(h^1).
\label{current fluctuations grid} 
\end{align}
Note that one could instead equivalently define the current with the normalization $h^{-2}$ to obtain densities and use the limit in Eq.~\eqref{limit grid}.

\begin{figure}
\includegraphics[width=1\textwidth]{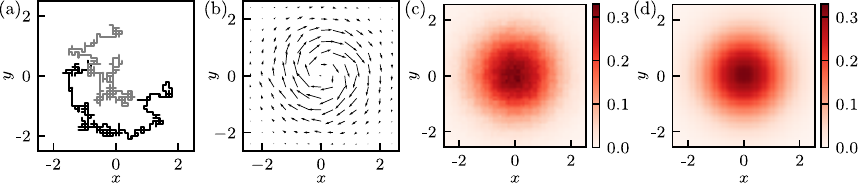}
\caption{(a) Two sample trajectories of the Ornstein-Uhlenbeck process discretized as described in the text with time step $\rmd t=0.001$ and total time $T=1$, starting in steady-state initial conditions. (b) Mean current as defined in the text obtained from a simulation of 10,000 trajectories such as the ones in (a). For visibility arrows are only drawn  at every fourth $x$ and $y$ value contained in the grid. (c) Variance of the current at individual grid points obtained from the same simulation as in (b). (d) Eq.~\eqref{current fluctuations grid} evaluated at individual grid points. The qualitative and quantitative agreement with (c) shows that Eq.~\eqref{current fluctuations grid}, which was derived in continuous space, has direct consequences for Markov jump processes on grids with small $h$. \label{FgMSMcurrfluc}}
\end{figure}

Fig.~\ref{FgMSMcurrfluc} illustrates the validity of the limit Eq.~\eqref{current fluctuations grid} for a discretized Ornstein-Uhlenbeck process, see Eq.~\eqref{OUP} with $r=D=1$, $\Omega=3$. The process on the area $[-5,5]^2$ is discretized into a $101\times 101=10201$-state Markov jump process, i.e.\ the grid spacing is $h=0.1$. The quantitative agreement of Fig.~\ref{FgMSMcurrfluc}c and d illustrates that the limit current fluctuations on the jump process are indeed governed by Eq.~\eqref{current fluctuations grid}. Note that the $t^{-1}$ scaling of the $h^{-2}$ term in Eq.~\eqref{current fluctuations grid} ensures that for $h\to 0$ current fluctuations are governed by this equation for all time-scales.

\section{Numerical and analytical evaluation used for the figures}\label{parameters and numerics} 
This section gives further parameters and all details necessary to reproduce all figures in the Letter. 

\subsection{Analytical results for the two-dimensional Ornstein-Uhlenbeck process}
For the numerical and analytical results shown in Figs.~1 and 3 in the Letter, we use the two-dimensional Ornstein-Uhlenbeck process given by the Langevin equation
\begin{align}
\rmd\x_t=\f F_{\rm rot}(\x_t)\rmd t+\sqrt{2D}\rmd\f W_t,\label{OUP}
\end{align}
with drift field $\f F_{\rm rot}(\x)=-\Theta\x$ where $\Theta=\begin{bmatrix}r&-\Omega\\\Omega&r\end{bmatrix},\ r>0$. The drift part splits into
\begin{align}
\f F_{\rm rot}(\x)=-D\{\nabla \phi(\x)\}+\js(\x)/\ps(\x),
\end{align}
with potential $\phi(\x)=\frac{r}{2D}\x^T\x$ and steady-state density and current 
\begin{align}
\ps(\x)&=\frac{r}{2\pi D}\rme^{-r\frac{\x^T\x}{2D}}\nonumber\\
\js(\x)&=(\f F-D\nabla_{\f x})\ps(\x)=\Omega\ps(\x)\begin{bmatrix}x_2\\-x_1\end{bmatrix}.\label{ps and js} 
\end{align}
A straightforward left-right decomposition \cite{SGardiner1985} gives the propagator/two-point function
\begin{align}
G(\x,t'|\x_0)&=\frac{r}{2\pi D(1-\rme^{-2rt'})}\exp\left[\frac{-r\left (\x-\rme^{-rt'}\begin{bmatrix}\cos(\Omega t')&\sin(\Omega t')\\-\sin(\Omega t')&\cos(\Omega t')\end{bmatrix}\x_0\right )^2}{2D(1-\rme^{-2rt'})\vphantom{\begin{bmatrix}0\\0\end{bmatrix}}}\right]\label{NESS-2d-OUP-propagator}
\nonumber\\
P_{\x_0}(\x,t')&\equiv G(\x,t'|\x_0)\ps(\x_0).
\end{align}
We then analytically solve the necessary Gaussian integrals for Gaussian window functions 
\begin{align}
U^h_\x(\z)=(2\pi h^2)^{-d/2}\exp\left[-\frac{(\z-\x)^2}{2h^2}\right],
\end{align}
and numerically solve the remaining $t'$-integral. This enables a very fast and stable (even for very small coarse graining where numerical spatial integrals would eventually fail) computation of the second moments shown in Figures 2 and 3. The analytical integrals were performed with the Python-based computer algebra system SymPy \cite{SSymPy}. To give an example, we now show the computation of one of the terms in the current variance result in Eq.~(5) in the Letter (other terms similarly).

We start e.g.\ with the spatial integrals
\begin{align}
\int\rmd^2x\int\rmd^2 x_0 V(\x)U(\x_0)j_{\rm s}^2(\x) j_{\rm s}^2(\x_0)G(\x,t|\x_0)\ps(\x_0),
\end{align}
where we set $\x=(x_1,x_2),\x_0=(x_3,x_4)$ such that $j_{\rm s}^2(\x)=-\Omega x_{1}$ and $j_{\rm s}^2(\x_0)=-\Omega x_{3}$ and we use constants $\{c_i\}$ to write
\begin{align}
V(\x)&=\frac{c_{1}}{\pi}\rme^{- c_{1} \left(\left(x_{1} - y_{1}\right)^{2} + \left(x_{2} - y_{2}\right)^{2}\right)}\nonumber\\
U(\x_0)&=\frac{c_{2}}{\pi}\rme^{- c_{2} \left(\left(x_{3} - y_{3}\right)^{2} + \left(x_{4} - y_{4}\right)^{2}\right)}\nonumber\\
\ps(\x_0)&=\frac{c_{3}}{\pi}\rme^{- c_{3} \left(x_{3}^{2} + x_{4}^{2}\right)}\nonumber\\
G(\x,t|\x_0)&=\frac{c_{4}}{\pi} \rme^{- c_{4} \left(\left(c_{5} x_{3} + c_{6} x_{4} - x_{1}\right)^{2} + \left(c_{5} x_{4} - c_{6} x_{3} - x_{2}\right)^{2}\right)}.
\end{align}
Integrating from $-\infty$ to $\infty$ over $x_3$ and $x_4$ gives (Gaussian integrals with $c_1,c_2,c_3,c_4>0$)
\begin{align}
&\int\rmd^2 x_0 V(\x)U(\x_0)j_{\rm s}^2(\x) j_{\rm s}^2(\x_0)G(\x,t|\x_0)\ps(\x_0)=\frac{\Omega^{2} c_{1} c_{2} c_{3} c_{4} x_{1} \left(c_{2} y_{3} + c_{4} c_{5} x_{1} - c_{4} c_{6} x_{2}\right)}{\pi^{3} \left(c_{2} + c_{3} + c_{4} c_{5}^{2} + c_{4} c_{6}^{2}\right)^{2}}\times\nonumber\\&
\rme^{\frac{c_{2}^{2} y_{4}^{2} + 2 c_{2} c_{4} y_{4} \left(c_{5} x_{2} y_{4} + c_{6} x_{1}\right) + c_{4}^{2} \left(c_{5} x_{2}+ c_{6} x_{1}\right )^{2}+ \left(c_{2} y_{3} + c_{4} c_{5} x_{1} - c_{4} c_{6} x_{2}\right)^{2}}{c_{2} + c_{3} + c_{4} c_{5}^{2} + c_{4} c_{6}^{2}}
- c_{1} x_{1}^{2} + 2 c_{1} x_{1} y_{1} - c_{1} x_{2}^{2} + 2 c_{1} x_{2} y_{2} - c_{1} (y_{1}^{2} + y_{2}^{2}) - c_{2} (y_{3}^{2} + y_{4}^{2}) - c_{4} (x_{1}^{2} + x_{2}^{2})}.
\label{supp1}
\end{align}
To integrate over $x_1$ and $x_2$, we simply use ($a_4,a_5>0$)
\begin{align}
\int_{-\infty}^\infty\rmd x_1&\int_{-\infty}^\infty\rmd x_2\left(a_{1} x_{1} + a_{2} x_{1}^{2} + a_{3} x_{1} x_{2}\right) \rme^{- a_{4} x_{1}^{2} - a_{5} x_{2}^{2} + a_{6} x_{1} + a_{7} x_{2} + a_{8}}
\nonumber\\&
=\frac{\pi \left(2 a_{1} a_{4} a_{5} a_{6} + a_{2} a_{5} \left(2 a_{4} + a_{6}^{2}\right) + a_{3} a_{4} a_{6} a_{7}\right) \rme^{\frac{4 a_{5} a_{8} + a_{7}^{2}}{4 a_{5}} + \frac{a_{6}^{2}}{4 a_{4}}}}{4 a_{4}^{\frac{5}{2}} a_{5}^{\frac{3}{2}}}.
\end{align}
Equations for the $\{a_i\}$ in terms of the $\{c_i\}$ can be read off Eq.~\eqref{supp1} and the $\{c_i\}$ contain all parameter dependencies of the process, including the $t'$. The $t'$-integration is then performed numerically. 

\subsection{Details and simulation parameters for figures in the Letter}
The process in Fig.~1a,b in the Letter is simulated as a free two-dimensional Brownian motion. Numerical Stratonovich integration gives the empirical density and current. The times shown are  $\tau_1^-=1.14$, $\tau_1^+=3.83$, $\tau_2^{-}=6.54$, $\tau_2^{+}=6.80$.

The mean and variance in the histograms in Fig.~1d,e and the relative error in Fig.~1f in the Letter are obtained analytically as described above. The TUR-bound is given by $\frac{2}{\sigma t}$ where $\sigma=\frac{2\Omega^2}{r}$ is the dissipation in the steady state of the Ornstein-Uhlenbeck process \eqref{OUP} (see \cite{SAccompanyingPaper}). For $\Omega=5,r=1,t=5$ we obtain that the TUR-bound shown in Fig.~1f in the Letter the is at $0.008$.

The process in Fig.~2 in the Letter is the shear flow with $\f F(x,y)=2x\hat\y$ and $\f D=\f 1$ from $(0,0)$ to $(2,0)$ in total time $t_2-t_1=1$. It is simulated with time step size $\rmd t=0.02$ as Brownian bridge in $x$-direction (exactly hits $2$ after time $1$) and then pick trajectories that hit $y_{\rm final}=0$ with deviation less than $0.02$. Time-reversed and dual reversed trajectories are similarly from $(2,0)$ to $(0,0)$ with same or inverted shear. For each transition around $11,000-12,000$ trajectories were considered. Arrows are in direction of the first/last step in discretized time.

The trajectory in Fig.~2e in the Letter is sampled from an Ornstein-Uhlenbeck process Eq.~\eqref{OUP} with $\Omega=3,r=D=1$ and total time $t\approx 37$. 

The simulations in Fig.~3 in the Letter are performed with time step $\rmd t=10^{-4}$ and $8192$ repetitions for 3a and $4096$ repetitions for 3b. All simulations are performed by discretizing Eq.~\eqref{OUP} and sampling the initial point $\x_0$ from the steady-state distribution $\ps(\x)$. Additional parameters in 3c are $h=1,0.25,0.03$ from dark to bright.

\end{document}